\documentclass[aps,prl,twocolumn,groupedaddress,nobibnotes]{revtex4-1}

\usepackage{graphicx}
\usepackage{caption}
\DeclareCaptionLabelSeparator{dot}{. }
\makeatletter
\def\justified{
	\let\\\@normalcr
	\@rightskip\z@skip \rightskip\@rightskip
	\leftskip\z@skip
	\parindent 0em\relax
	\setlength{\parfillskip}{0pt plus 1fil}}
\DeclareCaptionJustification{justified}{\justified}
\usepackage{subcaption}
\usepackage{amsmath}
\usepackage{braket}
\usepackage{units}
\usepackage{ragged2e}
\usepackage[colorlinks,urlcolor=blue ,citecolor=blue ,linkcolor=blue ]{hyperref}
\usepackage{xcolor}
\usepackage{textcomp}
\usepackage{bm}
\usepackage{dsfont}
\usepackage{physics}
\usepackage{titlesec}
\usepackage{blkarray}
\titlespacing*{\section}{0pt}{3.5ex plus 2ex minus .2ex}{2.3ex plus .2ex}

\newcommand\xleftrightarrow[1]{%
	\mathbin{\ooalign{$\,\xrightarrow{#1}$\cr$\xleftarrow{\hphantom{#1}}\,$}}
}

\begin{document}
	
	
	\title{Orbital and Spin Dynamics of Single Neutrally-Charged Nitrogen-Vacancy Centers in Diamond}
	
	\author{S. Baier$^{1,2,\dagger}$}
	\thanks{These authors contributed equally to this work.}
	\affiliation{%
		$^{1}$QuTech, Delft University of Technology, 2628 CJ Delft, The Netherlands\\
		$^{2}$Kavli Institute of Nanoscience, Delft University of Technology, 2628 CJ Delft, The Netherlands
	}
	\author{C. E. Bradley$^{1,2}$}
	\thanks{These authors contributed equally to this work.}
	\affiliation{%
		$^{1}$QuTech, Delft University of Technology, 2628 CJ Delft, The Netherlands\\
		$^{2}$Kavli Institute of Nanoscience, Delft University of Technology, 2628 CJ Delft, The Netherlands
	}
	\author{T. Middelburg$^{1,2}$, V. V. Dobrovitski$^{1,2}$, T. H. Taminiau$^{1,2}$}
	\affiliation{%
		$^{1}$QuTech, Delft University of Technology, 2628 CJ Delft, The Netherlands\\
		$^{2}$Kavli Institute of Nanoscience, Delft University of Technology, 2628 CJ Delft, The Netherlands
	}
	\author{R. Hanson$^{1,2,\dagger}$}
	\affiliation{%
		$^{1}$QuTech, Delft University of Technology, 2628 CJ Delft, The Netherlands\\
		$^{2}$Kavli Institute of Nanoscience, Delft University of Technology, 2628 CJ Delft, The Netherlands
	}


	\date{\today}
	
	\begin{abstract}{
			The neutral charge state plays an important role in quantum information and sensing applications based on nitrogen-vacancy centers. However, the orbital and spin dynamics remain unexplored. Here, we use resonant excitation of single centers to directly reveal the fine structure, enabling selective addressing of spin-orbit states. Through pump-probe experiments, we find the orbital relaxation time (430~ns at 4.7~K) and measure its temperature-dependence up to 11.8~K. Finally we reveal the spin relaxation time (1.5~s), and realize projective high-fidelity single-shot readout of the spin state ($\geq98\%$).
		}
	\end{abstract}
	
	\maketitle
	
	
	Defect centers in solids are a promising class of systems for quantum science and technology \cite{atature2018material,awschalom_quantum_2018}. They combine bright optical transitions, access to long-lived electronic- and nuclear-spin registers and compatibility with solid-state device engineering. Of particular prominence is the negatively-charged nitrogen-vacancy center (NV$^{-}$) in diamond, which has 
	enabled recent advances in quantum information science \cite{humphreys_deterministic_2018,bradley2019ten} and quantum sensing \cite{aslam2017nanoscale,glenn2018high,abobeih2019atomic}.
	
	Alongside NV$^{-}$, the nitrogen-vacancy defect can exist in both the neutral- (NV$^{0}$) and --- with sufficient Fermi-level engineering --- positive- (NV$^{+}$) charge states. 
	These additional charge states can be used as a resource in a number of applications, such as spin-to-charge conversion for improved spin-state read-out~\cite{shields2015efficient,hopper2018spin}, classical data storage in NV ensembles \cite{dhomkar2016long}, and deliberate charge-state switching for improved nuclear-spin coherence under ambient conditions \cite{maurer_room-temperature_2012,pfender_protecting_2017}.
	
	Conversely, for experiments based upon NV$^{-}$, undesired conversion to NV$^{0}$ can be a hindrance: active charge-state initialization protocols have been used to counter this \cite{bernien_heralded_2013,hopper2020real}. For quantum networks, stochastic conversion from NV$^{-}$ to NV$^{0}$ is an important decoherence mechanism for nuclear-spin quantum memories \cite{kalb_dephasing_2018}.
	
	Despite the importance of NV$^{0}$, understanding of many of its properties remains elusive. In particular, the orbital- and spin-dynamic timescales are unknown. Also, while recent magnetic circular dichroism (MCD) measurements on ensembles \cite{barson_fine_2019,braukmann_circularly_2018} give insight into the NV$^0$ fine structure, no direct observation has been reported. Building an understanding of the system and its associated dynamic processes is important for improving control in NV quantum devices. Moreover, the knowledge gained may offer new insights into the physics of other impurities in solids \cite{bassett2019quantum}. Finally, NV$^{0}$ may prove to be a powerful quantum system in its own right.
	
\begin{figure}
		\includegraphics[width=1\linewidth]{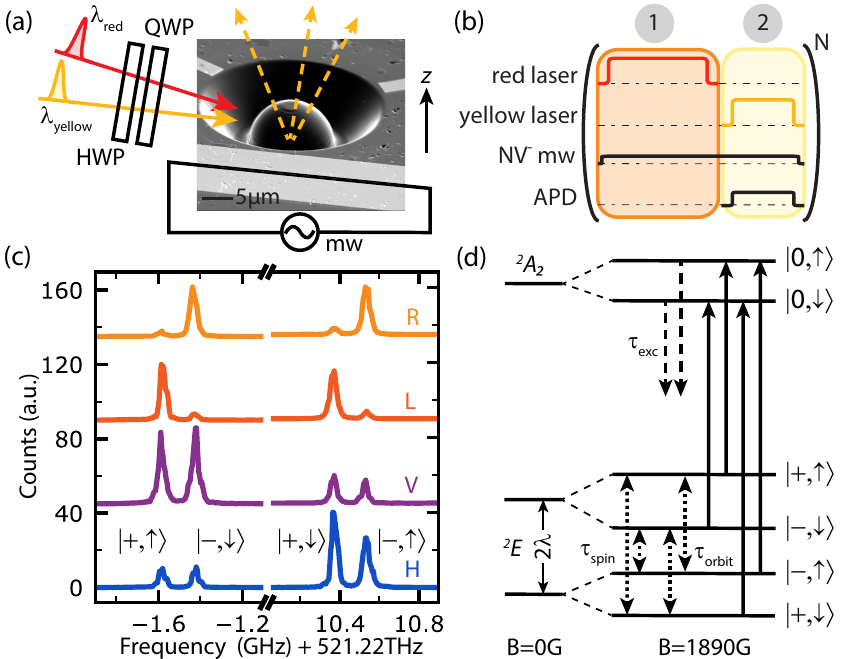}%
		\caption{\label{fig:levels} Direct observation of the fine structure of the NV$^{0}$ center. (a) Electron microscope image of a solid immersion lens fabricated around the NV center. Optical ($\lambda_{\rm yellow}$, $\lambda_{\rm red}$) and mw control are indicated. (b) Experimental sequence for spectroscopy consisting of a preparation (1) and measurement (2) part. (c) Spectra obtained with linear (H,V) and circular (L,R) polarizations ($P_{\rm yellow}=500\,\mathrm{pW}$), offset for clarity~\cite{supp}. (d) Ground and excited state level structure. Spin-conserving optical transitions (solid arrows), excited state decay (dashed arrows) and spin/orbital relaxations (dotted arrows) are indicated.}
	\end{figure}

	Here, we develop protocols combining resonant excitation of both NV$^{0}$ and NV$^{-}$. We apply these novel protocols to reveal the orbital and spin dynamics of single NV$^{0}$ centers in diamond, as well as to realize initialization and single-shot readout of the NV$^{0}$ spin state. We perform our measurements on single NV centers at cryogenic temperatures~\cite{supp}, see Fig.\,\ref{fig:levels}(a). The NV center is adressed with microwave (mw) pulses (NV$^-$ ground-state spin transitions) as well as with polarization-controlled $\lambda_{\rm red}=637\,\mathrm{nm}$ (NV$^-$ zero-phonon line (ZPL)) and $\lambda_{\rm yellow}=575\,\mathrm{nm}$ (NV$^{0}$ ZPL) laser light. We apply an axial magnetic field of $B_z=1890(5)\,\mathrm{G}$ to induce significant Zeeman splitting.
	
	The ZPL of the NV$^{0}$ center has been conclusively attributed to this defect \cite{davies_dynamic_1979,mita_change_1996,kennedy2003long,gaebel2006photochromism,waldherr_dark_2011,aslam_photo-induced_2013}. A combination of ab-initio calculations and symmetry arguments led to the proposal of ground states of $^{2}$E symmetry, which can be optically excited to a $^{2}$A$_{2}$ manifold \cite{gali_theory_2009,manson_assignment_2013}. An additional metastable $^{4}$A$_{2}$ quartet state was also predicted, and has been observed by electron paramagnetic resonance (EPR) measurements under excitation of the NV$^{0}$ ZPL \cite{felton_electron_2008}. A splitting of the transitions of the two orbital states $E_x$ and $E_y$ has been measured~\cite{manson_assignment_2013,hensen_quantum_2016}. However, the associated fine structure has not been observed in PL or EPR measurements. 
	
	We start by performing spectroscopy using the experimental procedure sketched in Fig.\,\ref{fig:levels}(b). For each frequency step, we (1) probabilistically prepare the emitter in NV$^{0}$ by applying strong laser excitation resonant with the NV$^{-}$ ZPL, in combination with weak mw driving~\cite{supp} to induce the conversion NV$^{-}$ $\rightarrow$ NV$^{0}$. We then (2) apply polarized yellow light, during which time all single-photons above $650$\,nm are integrated.
	
	The measured spectra (Fig.\,\ref{fig:levels}(c)) show four transitions --- the first direct spectroscopic observation of the NV$^{0}$ fine structure. These observations validate the model of Barson et al. \cite{barson_fine_2019}, and we hence follow their theoretical description below. Under the secular approximation, the ground-state Hamiltonian of NV$^{0}$ can be described by 
	\begin{equation}
	\label{eq:hamiltonian}
	\begin{split}
	H  = & \: g\mu_{B}\hat{S}_{z}B_{z} + l\mu_{B}\hat{L}_{z}B_{z} + 2 \lambda \hat{L}_{z}\hat{S}_{z}\\ 
	& + \epsilon_{\bot} (\hat{L}_{-} + \hat{L}_{+}).
	\end{split}
	\end{equation}
	$g$ is the spin g-factor, $\mu_{B}$ is the Bohr magneton, $l$ is the orbital g-factor, $\lambda$ is the spin-orbit interaction parameter and $\epsilon_{\bot}$ is the perpendicular strain parameter. $\hat{L}_{z} = \sigma_{z}$ and $\hat{S}_{z} = \frac{1}{2}\sigma_{z}$ are the orbital and spin operators defined in terms of the Pauli matrix $\sigma_{z}$, while $\hat{L}_{\pm}= \ket{\pm}\bra{\mp}$ with $\ket{\pm} = \mp(1/\sqrt{2}(\ket{X}\pm i\ket{Y}))$ are the orbital operators defined within the basis of the strain eigenstates \{$\ket{X},\ket{Y}$\}. The $z$-axis is defined parallel to the NV axis.
	
	The resulting level structure is presented in Fig.\,\ref{fig:levels}(d). The $^{2}$E ground state is composed of a pair of doublet states with opposite spin-orbit parity (lower spin-orbit branch: $\{\ket{+,\downarrow}, \ket{-,\uparrow}\}$; upper spin-orbit branch: $\{\ket{-,\downarrow}, \ket{+,\uparrow}\}$). The degeneracy of each doublet is lifted by orbital- and spin-Zeeman contributions under the applied magnetic field. Conversely, the $^{2}$A$_{2}$ excited state exhibits no spin-orbit structure, but is rather split by the spin-Zeeman effect alone.
	These contributions lead to four spin-conserving transitions. The contributing ground state for each observed transition is indicated in Fig.\,\ref{fig:levels}(c).
	
	We find that the luminescence of the transitions depend significantly on the polarization of the excitation light (see Fig.\,\ref{fig:levels}(c)). Differing transition amplitudes for orthogonal polarizations can be attributed to optical selection rules that are strongly dependent on $\epsilon_{\bot}$~\cite{supp,barson_fine_2019}.
	Based upon these observations, we develop a method to extract $\epsilon_{\bot}$ and simultaneously the fine structure parameters of the NV$^0$ Hamiltonian~\cite{supp}. By fitting spectra from three individual NV centers against our theoretical model, we find $l=0.039(11)$ and $\lambda=4.9(4)\,\mathrm{GHz}$. These values are roughly a factor of 2 larger than those found previously using NV-ensemble MCD measurements~{\footnote{A re-assessment of the procedures of Barson et al.\, concluded that it can not be excluded that an error in documenting the data is the cause of this discrepancy~\cite{private_comm}}}.
	
	Crucially, the data in Fig.\,\ref{fig:levels}(c) shows that resonant optical excitation in this magnetic field regime allows for state-resolved addressing, enabling the heralded preparation of specific states and investigation of the system dynamics. To date, only the excited-state lifetime, $\tau_{\text{exc}}$, of 21 ns has been reported \cite{beha_optimum_2012}. Here, we investigate the orbital- and spin-relaxation timescales of the ground state, $\tau_{\text{orbit}}$ and $\tau_{\text{spin}}$, see Fig.\,\ref{fig:levels}(d).
	
	\begin{figure}
		\includegraphics[width=1\linewidth]{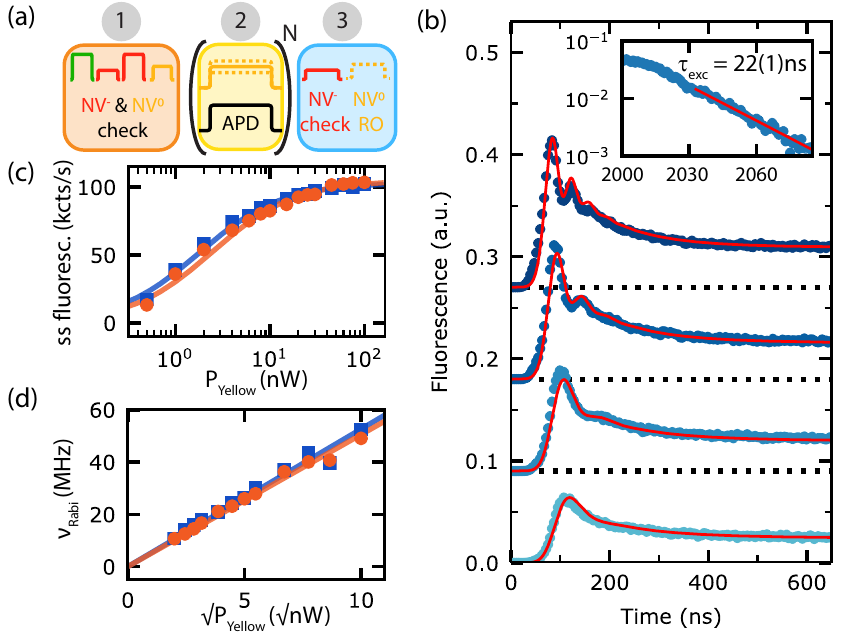}%
		\caption{\label{fig:drive} Time-resolved resonant pump measurements. (a) Experimental sequence consisting of preparation (1), measurement (2) and readout (3) parts. (b) Fluorescence of NV$^{0}$ when driving the lower spin-orbit branch with H polarisation for $P_{\rm yellow}=2,4,10,20\,\mathrm{nW}$ (bottom to top) averaged over at least $1\times10^{6}$ repetitions. Measurements have a timing resolution of $250\,\mathrm{ps}$ and are offset for clarity. Solid red lines are simulations of the full system dynamics with our theoretical model~\cite{supp}. Inset: decay of fluorescence counts after the AOM is closed. (c) Steady state (ss) fluorescence counts as a function of $P_{\rm yellow}$, for H (squares) and L polarisation (circles). The data is fit with a saturation curve $f(P) =  A (P/(P + P_{\text{sat}}))$. (d) Optical Rabi frequency as a function of $\sqrt{P_{\rm yellow}}$. Fits yield a slope of 5.3(1)/5.1(1) $\mathrm{MHz}/\sqrt{\mathrm{nW}}$ for L/H polarization.}
	\end{figure}
	
	In order to unambiguously measure the dynamics of NV$^{0}$, we design and implement a charge-resonance (CR) protocol that realizes high-fidelity heralded preparation into NV$^{0}$, with the $\lambda=575\,\mathrm{nm}$ laser resonant with a chosen optical transition, see Fig.\,\ref{fig:drive}(a). The CR protocol (1) can be broken down as follows. First, a heralding signal confirms preparation in NV$^{-}$, with the $\lambda=637\,\mathrm{nm}$ lasers on resonance with the NV$^{-}$ transitions. Next, a strong red optical pulse induces charge state conversion, after which a chosen NV$^{0}$ transition is excited with yellow light. If the photon counts obtained during the `NV$^{0}$ check' exceed a pre-set threshold, the protocol is completed. Further details are given in the supplementary information \cite{supp}. 
	
	After the CR protocol, we perform the experimental sequence on NV$^{0}$ (2). Finally, we detect whether undesired conversion to NV$^{-}$ occurred during the experimental sequence, and then perform read-out of the NV$^{0}$ state (3). Note that the CR protocol prepares a specific spin state of the NV$^{0}$ center. For circular polarisation we typically start the experiment by heralding the $\ket{\downarrow}$ spin state. For linear polarisation, however, due to their close spectral vicinity, the CR check heralds either the $\ket{\downarrow}$ or $\ket{\uparrow}$ spin state.
	
	In Fig.\,\ref{fig:drive}(b) we show time-resolved pump measurements. Here, the yellow laser is gated by an acousto-optic modulator (AOM), with measured rise-/fall-time of $30(5)$/$7(1)\,\mathrm{ns}$. Upon opening the AOM, we observe a rapid increase in fluorescence due to optical cycling, which is then damped as population is pumped out of the driven state. By fitting the steady-state fluorescence counts for L/H polarization we extract a saturation power of $2.5(2)$/$1.8(1)\,\mathrm{nW}$ and saturation counts of $105(2)$/$103(2)\,\mathrm{kcts/s}$, see Fig.\,\ref{fig:drive}(c). As the optical power is increased, coherent optical Rabi oscillations are observed. In Fig.\,\ref{fig:drive}(d), we plot the fitted frequency of these oscillations, revealing the expected $\sqrt{P_{\rm yellow}}$ dependence. When the AOM is closed the fluorescence decays with $\tau_{\text{exc}} = 22(1)\,\mathrm{ns}$ (inset Fig.\,\ref{fig:drive}(b)), which is consistent with literature~\cite{beha_optimum_2012}.
	
	\begin{figure}
		\includegraphics[width=1\linewidth]{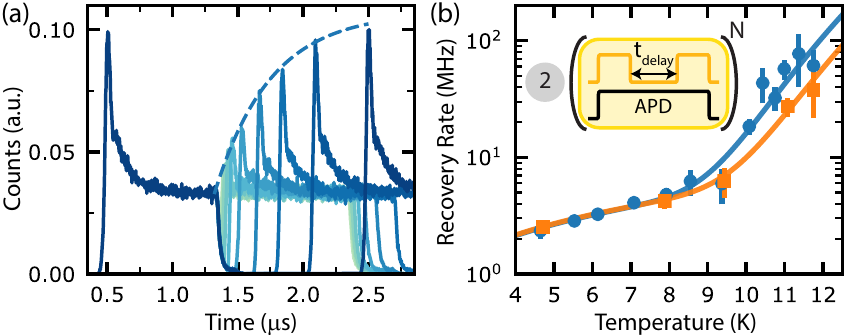}%
		\caption{\label{fig:orbit} Time-resolved pump-probe spectroscopy. The experimental sequence after state preparation is given in the inset of (b). (a) Example traces for a range of $t_{\rm delay}$ (light to dark for increasing $t_{\rm delay}$), at a temperature of 5.5(1) K, integrated over $5\times10^{6}$ acquisitions each, measured with H polarisation. The dashed line is a fit to the recovery behaviour \cite{supp}. (b) Recovery rate $R_{\mathrm{recovery}}$ as a function of the cryostat temperature. Circles (squares) describe data measured on the lower (upper) spin-orbit branch. Error bars for $R_{\mathrm{recovery}}$ correspond to 1 s.\,d.\, fit errors. The solid lines are fits of form $f(T) = A \ T + B \ \mathrm{exp}[-\Delta/k_B T]$, giving $A=0.53(3)\,$MHz/K ($A=0.54(2)\,$MHz/K) and $B=1(1)\times10^7\,$MHz ($B=1(4)\times10^7\,$MHz) for the lower (upper) branch.}
	\end{figure}
	
	To uncover the recovery timescale after pumping we turn to pump-probe spectroscopy. Example time-traces are shown in Fig.\,\ref{fig:orbit}(a). The resulting data is well described by an exponential recovery with a single timescale associated with how fast the system relaxes~\cite{supp} once illumination is turned off. At base temperature of our cryostat ($T=4.65(3)\,\mathrm{K}$), we extract $\tau_{\mathrm{recovery}}=0.43(6)\,\mathrm{\mu s}$. We  attribute these fast dynamics to orbital relaxation processes, i.e. $\ket{+}\leftrightarrow\ket{-}$ and $\tau_{\mathrm{orbit}}=\tau_{\mathrm{recovery}}$.
	
	We repeat the pump-probe measurements across a range of temperatures. The fitted recovery times are shown as rates $R_{\rm recovery}=1/t_{\rm recovery}$ in Fig.\,\ref{fig:orbit}(b). After an initial linear increase a rapid increase is observed at higher temperatures. 
	At these higher temperatures, the required time resolution exceeds the AOM switching time constants, which we take into account in the fitting procedure~\cite{supp}.
	
	The initial linear increase ($\propto T$) can be attributed to single-phonon processes, while high-order processes appear to govern the recovery rate at higher temperatures \cite{orbach1961spin,abragam2012electron}. Here, we fit individually to a two-phonon Raman process ($\propto T^n$) and a two-phonon Orbach process ($\propto \mathrm{exp}[-\Delta/k_B T]$), with $k_B$ being the Boltzmann constant. For the Raman process the fit returns $n=13(2)$ ($14(3)$) for the lower (upper) spin-orbit branch; a physical explanation for such values is currently lacking. For the Orbach process we find a characteristic energy scale of $\Delta=12(2)\,\mathrm{meV}$ ($\Delta=13(4)\,\mathrm{meV}$) extracted from a fit to the lower (upper) spin-orbit branch. $\Delta$ is associated to the energy splitting to the first vibronic level of the NV$^{0}$ ground state, predicted to be a Jahn-Teller system~\cite{davies_dynamic_1979,zhang2018multimode}. The value found here agrees with the bulk absorption measurements of Davies~\cite{davies_dynamic_1979} ($13.6(7)\,\mathrm{meV}$), and with recent density-functional theory calculations ($21.4\,\mathrm{meV}$)~\cite{zhang2018multimode}, suggesting that the measured increase of $R_{\rm recovery}$ is predominantly due to two-phonon Orbach processes. While a detailed model is beyond the scope of this work, we expect that our findings will aid in the further understanding of the vibronic structure of NV$^{0}$.

	\begin{figure}
		\includegraphics[width=1\linewidth]{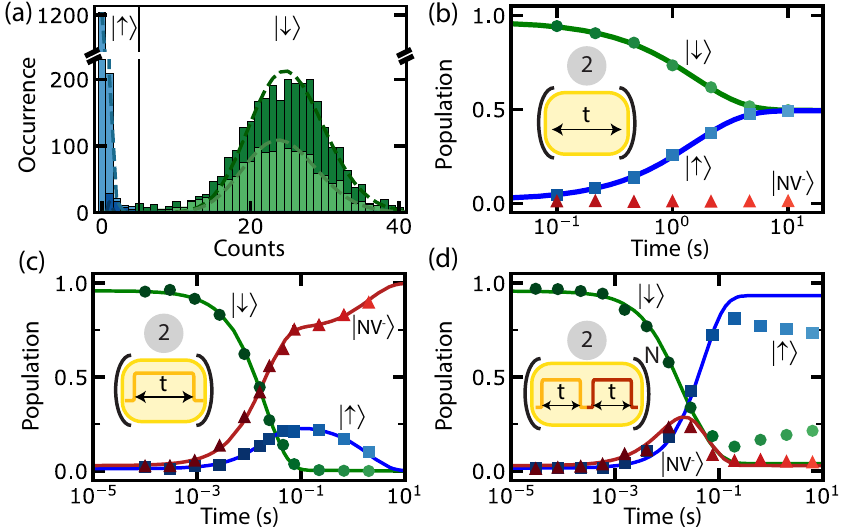}%
		\caption{\label{fig:spin} Single-shot readout and spin pumping. (a) Histograms after preparation of the NV$^{0}$ $\ket{\downarrow}$ state (dark colors) and mixed (light colors) spin states. (b) Relaxation (T$_{1}$) measurement for the $\ket{\downarrow}$ (circles) and $\ket{\uparrow}$ (squares) states, fitted with an exponential decay (recovery). The NV$^{-}$ population (triangles) remains negligible in the dark. The data is averaged over $3\times10^3$ repetitions each. (c) Spin-pumping: NV$^{0}$ spin- and total NV$^{-}$ populations as a function of yellow illumination time. Solid lines are fits to solutions for the underlying 3-level rate equations \cite{supp}.  (d) Spin-pumping with charge-cycling: same as (c) but with stroboscopic red illumination. The time axis is the yellow illumination time (half of the total sequence time).}
	\end{figure}
	
	Now we turn to the spin dynamics of NV$^0$. Here, we exploit polarization control to selectively prepare, address, and readout the NV$^{0}$ spin state. These measurements are all performed on timescales $\gg$ $\tau_{\mathrm{orbit}}=0.43(6)\,\mathrm{\mu s}$ and thus average over the orbital basis; we will therefore only refer to the spin  states. In all experiments below, we use L polarisation, addressing the $\ket{\downarrow}$ state. We herald the preparation of $\ket{\downarrow}$ by applying 25 nW for $250\,\mathrm{\mu s}$, and proceed when more than 25 photons are detected. After a delay of $0.1\,\mathrm{ms}$, we perform a charge-state check with red excitation, followed by a second yellow readout, see Fig.\,\ref{fig:drive}(a)(3). We then repeat this experiment, but with a delay of $10\,\mathrm{s}$ between the yellow readouts, allowing for relaxation processes to occur. The resulting histograms are shown in Fig.\,\ref{fig:spin}(a).
	
	In the first case (dark colors), we observe a single dominant population which can be modelled by a Poissonian distribution with mean photon count 25.2(2), and that we attribute to $\ket{\downarrow}$. In the second case (light colors), we additionally observe a second distribution, again modelled as a Poissonian distribution with mean photon count 0.171(4). A charge-state measurement of NV$^{-}$ performed before each read-out shows that only a small fraction of the population ($P_{\rm NV^-}\sim$ 1\%) is found in the unwanted charge state --- which we discard from the histograms --- and that the majority of low-count events can be attributed to a dark state of NV$^{0}$. 
	As the populations evolve without laser excitation, the dark state must be part of the ground state manifold; we therefore assign this state to the second spin state $\ket{\uparrow}$.
	A read-out threshold of 5 photons (solid line, Fig.\,\ref{fig:spin}(a)) discriminates the two spin states. 
	
	We now sweep the delay time between initialisation and read-out. The measured populations of $\ket{\downarrow}$ ($P_{\downarrow}$) and $\ket{\uparrow}$ ($P_{\uparrow}$) are plotted in Fig.\,\ref{fig:spin}(b), showing relaxation to a mean population of 0.494(6). The data is consistent with a spin-1/2 T$_{\text{1}}$ process of characteristic timescale $\tau_{\mathrm{spin}}=1.51(1)\,\mathrm{s}$. Note that the observed value is a lower bound of the intrinsic spin relaxation, as it may be limited by leakage of resonant laser light. By setting the initial and long-time population in $\ket{\downarrow}$ to be 1 and 0.5 respectively, we obtain a lower-bound for the single-shot read-out fidelity, $F_{RO} = \frac{1}{2}(F_{\ket{\downarrow}} + F_{\ket{\uparrow}}) \geq 98.2(9)\%$, where $F_{\ket{s}}$ is the probability to assign $\ket{s}$ after preparing $\ket{s}$~\cite{supp}. 
	
	To investigate the cycling nature of the driven optical transition we now repeat the measurement under $5\,\mathrm{nW}$ of resonant yellow excitation, see Fig.\,\ref{fig:spin}(c). We find that $P_{\downarrow}$ decreases on a timescale faster than can be explained by spin-relaxation alone, showing that the optical excitation induces spin pumping. Possible spin-mixing channels are given either in the $^2A_2$ excited state or via an intersystem crossing, which might be offered by the $^{4}A_{2}$ state. 
	We also find a significant increase of $P_{\rm NV^-}$ due to optically-induced charge conversion \cite{siyushev_optically_2013, supp}. However, this slows once $\ket{\downarrow}$ is depleted as $\ket{\uparrow}$ is a dark state for optical excitation. Beyond this, $P_{\uparrow}$ reduces with $\tau_{\mathrm{spin}}$, and charge conversion continues. We find a high state preparation fidelity for $\ket{\uparrow}$ of $99^{+1}_{-10}\,\%$ after $600\,\mathrm{ms}$, but with an absolute population in the NV$^{0}$ $\ket{\uparrow}$ state of only $22(2)\,\%$.
	
	To reveal the respective rates we develop a three-level-rate equation model that we fit to our data, using the measured spin-relaxation time as a fixed input (solid lines, Fig.\,\ref{fig:spin}(c)) \cite{supp}. For the applied power of 5 nW, we extract characteristic timescales of $27(1)\,\mathrm{ms}$ ($90(4)\,\mathrm{ms}$) for the charge conversion (spin pumping) process. From this we can estimate the cyclicity of the $\ket{\downarrow}$ state within this regime to be $0.98(8)\times 10^{5}$ cycles, mainly limited by recharging to NV$^{-}$~\cite{supp}. 
	
	In a second experiment the $5\,\mathrm{nW}$ yellow excitation is stroboscopically interleaved with strong NV$^{-}\rightarrow{}$NV$^{0}$ ionisation pulses~\cite{supp}, see Fig.\,\ref{fig:spin}(d). Again we observe a gradual decrease of $P_{\downarrow}$, and an increase of both $P_{\uparrow}$ and $P_{\mathrm{NV}^-}$, but then $P_{\mathrm{NV}^-}$ growth stops and even inverts. This observation can be explained via the picture that the removal of an electron from NV$^{-}$ prepares a random spin-state in NV$^{0}$, eventually populating the dark state $\ket{\uparrow}$. Competing rates between this spin-selection process and spin relaxation lead to the observed steady state populations.
	We again fit a three-level rate equation model, using the previously obtained parameters as fixed inputs \cite{supp}, and extract a timescale for ionisation of $18(4)\,\mathrm{ms}$. The rate equation model does not accurately describe the behaviour at long timescales, which is likely due to a reduction of the NV$^{0}$ spin-relaxation time under red excitation and strong NV$^{-}$ microwave driving~\cite{supp}.
	
	As a final step, we develop a master equation simulation to capture the full dynamics of the NV$^0$ center~\cite{supp}. In Fig.\,\ref{fig:drive}(b) we plot the simulated excited state population (solid line), using the uncovered NV$^0$ timescales and spectral properties. We match the Rabi frequency to the measured optical power and further include a spectral average over a Gaussian distribution of detuning values with FWHM $=2\pi\times 20\,\mathrm{MHz}$. We find excellent agreement with our experimental fluorescence data, emphasizing a consistent understanding of the NV$^0$ dynamics.
	
	In conclusion, we have developed a novel toolbox for the study and control of single neutrally-charged NV centers in diamond. We have uncovered the dynamic timescales and demonstrated single-shot readout and initialization-by-measurement of the NV$^{0}$ spin, each with high fidelity. In future investigations, coherent control of the spin states may be obtained. Detailed modelling of the defect may give new insights into the observed temperature-dependence of the orbital dynamics. On the application side, protection of nuclear spin quantum memories from dephasing by NV$^{0}$ may be achieved by microwave spin locking in both orbitals, or by feedback based upon the NV$^{0}$ spin read-out demonstrated here. Finally, at reduced temperatures that suppress the orbital dynamics, NV$^{0}$ may prove to be a powerful system for quantum technologies in its own right.
	
	\begin{acknowledgments}
		We thank Michael Barson, Marcus Doherty and Neil Manson for fruitful discussions. Further, we thank Matteo Pompili, Sophie Hermans and Hans Beukers for experimental assistance, and Joe Randall, Maximilian Ruf and Matteo Pasini for reviewing the manuscript. 
		We acknowledge financial support from the EU Flagship on Quantum Technologies project Quantum Internet Alliance, from the Netherlands Organisation for Scientific Research (NWO) through a VICI grant, a VIDI grant (Project No.\,680-47-552) and within the research programme NWO QuTech Physics Funding (QTECH, programme 172, Project No.\,16QTECH02), and the Zwaartekracht Grant Quantum Software Consortium (Project No.\,024.003.037/3368), and the European Research Council (ERC) through an ERC Consolidator Grant and an ERC Starting Grant (grant agreement No. 852410). S.B is supported within an Erwin-Schr\"odinger fellowship (QuantNet, No.\,J\,4229-N27) of the Austrian National Science Foundation (FWF).
		
		$^\dagger$ Correspondence and requests for materials should be addressed to s.baier@tudelft.nl or r.hanson@tudelft.nl
	\end{acknowledgments}
	

%

	\appendix
	\renewcommand\thefigure{\thesection S\arabic{figure}}   
	\setcounter{figure}{0}   
	\section{Supplemental Material}
	\subsection{Experimental setup} 
	
	Our experiments are performed on single nitrogen-vacancy (NV) centers in type-IIa bulk diamond (Element Six, CVD grown, $<$111$>$ oriented), using a cryogenic (Montana Cryostation, $4\,\mathrm{K}$) home-built confocal microscope setup. Enhanced photon-collection efficiency is achieved by fabrication of solid immersion lenses \cite{hadden2010strongly} and an anti-reflection coating~\cite{bernien_control_2014}. 
	For phonon-sideband (PSB) detection, a dichroic mirror (Semrock, pass above $650\,\mathrm{nm}$) and an additional long-pass filter are used to block reflections of the excitation lasers. Photon emission is detected via an avalanche photo-diode (APD, Laser components, quantum efficiency $\sim 80\%$), with a total collection efficiency of $\sim 3\%$ ($\sim 10\%$) of the  NV$^0$ (NV$^-$) PSB. 
	
\begin{figure}
		\includegraphics[width=1\linewidth]{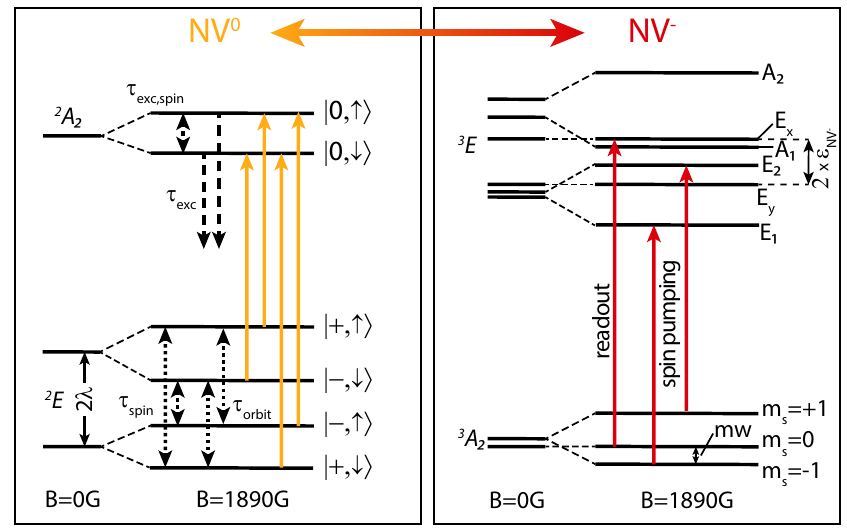}%
		\caption{\label{fig:levels_supp} Level structure for the two charge states NV$^0$ and NV$^-$ of a single NV center. The optical transitions used within this work are indicated by yellow (red) solid arrows for NV$^0$ (NV$^-$). Charge-state switching between the two charge states is achieved via two-photon absorption of the respective ZPL laser~\cite{siyushev_optically_2013}.}
	\end{figure}

	We apply a magnetic field of $B_{z}=1890(5)\,\mathrm{G}$ along the symmetry axis $z$ of the NV center via a permanent magnet. A slight misalignment of the field leaves a small perpendicular magnetic field component of $B_{\bot}=10(5)\,\mathrm{G}$. As the ratio $B_{\bot}/B_{z}$ is small, we neglect the effect of the perpendicular magnetic field.
	In addition to the magnetic field, local strain and electric fields can alter the NV center level-structure~\cite{doherty_nitrogen-vacancy_2013}. For the NV center used in the main text we observe the level structures as depicted in Fig.\,\ref{fig:levels_supp}. 
	
	To address the NV$^0$ charge state we apply resonant optical excitation of the NV$^0$ zero-phonon-line (ZPL) ($\lambda = 575.17\,\mathrm{nm}, \omega=2\pi\times521.22\,\mathrm{THz}$). The laser frequency can be manually tuned to each $^2E$ to $^2A_2$ transition.
	In the NV$^-$ charge state, selective excitation of ZPL transitions ($\lambda = 637.25\,\mathrm{nm}, \omega=2\pi\times470.45\,\mathrm{THz}$) enables optical readout (RO, $m_{\rm s}= 0 \xleftrightarrow{} E_{\rm x}$) and spin-pumping (SP, $m_{\rm s}= \pm 1 \xleftrightarrow{} E_{1,2}$).
	The ground state spin levels ($m_s=0\xleftrightarrow{} m_s=-1$) can be coherently adressed with microwave (mw) pulses delivered via gold strip lines on top of the diamond surface.
	For the NV used in the main text (NV A), we extract an NV$^{-}$ perpendicular strain of $\epsilon_{\rm NV^-}=4.2(1)\,\mathrm{GHz}$, from the observed optical transition frequencies, see Fig.\,\ref{fig:levels_supp}. We note that it is unclear how this relates to the strain in NV$^0$, as the susceptibility of the NV$^0$ states to electric fields is currently unknown. Further, charge state conversion may result in differing local charge environments for the two charge states~\cite{manson_nv-n_2018}.
	
	\subsection{Photo-luminescence measurements} 
	
	For the photo-luminescence (PL) measurements presented in Fig.\,1 of the main text, the following methodology was used. First, the NV center is prepared in the neutral charge state by the application of a $50\,\mathrm{\mu s}$ optical SP pulse ($1\,\rm \mu W$) in combination with weak mw driving ($\nu_{\rm Rabi}^{\rm mw}\sim 150\,\rm kHz$) of the NV$^{-}$ $m_{s} = 0 \xleftrightarrow{} m_{s} = -1$ ground-state spin transition.
	This method prevents optical spin-pumping into a NV$^{-}$ dark state ($m_{s} = 0$). Second, $500\,\mathrm{pW}$ of yellow light (${P_{\rm yellow}}$) is applied for $15\,\mathrm{\mu s}$ at the NV$^{0}$ ZPL transition, during which all single-photon detection is integrated. The red-yellow procedure is repeated $N=150$ times before the frequency is stepped by $1\,\mathrm{MHz}$.
	
	A total of 20 full scans were made for each polarization setting, which were collated to produce the final PL data. For each PL scan, the fluorescence maxima are found via a peak finding routine (python, scipy.signal.find\_peaks\_cwt). Further, all data sets are shifted to the mean frequency of all fluorescence maxima and summed. We typically observe shifts of the maxima by up to $200\,\mathrm{MHz}$ due to spectral diffusion. PL scans of the lower and upper spin-orbit branches were done in two separate measurements. To avoid systematic shifts of the splitting between the two spin-orbit branches, the NV center is reset by strong green ($\lambda=515\,\mathrm{nm}$, $10\,\rm \mu W$) illumination in between these two measurements, cancelling potentially accumulating effects of spectral diffusion from the red-yellow scans.
	
\begin{figure}
		\includegraphics[width=1\linewidth]{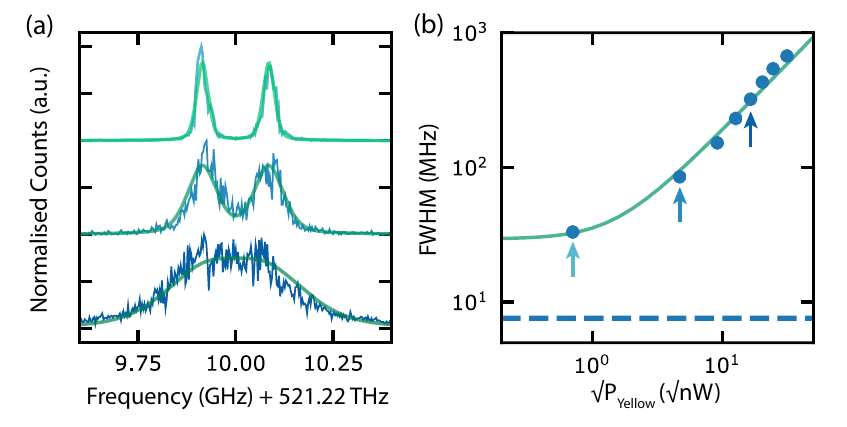}%
		\caption{\label{fig:broadening} Power broadening of NV$^{0}$ optical transitions. (a) Example PL spectra for the lower spin-orbit branch, measured at powers of 0.5 nW, 22 nW, and 160 nW (light to dark blue) with H polarisation. Each spectrum consists of 20 individual scans which have been shifted on top of one another and summed. A pair of Voigt profiles is fitted. The spectra have been normalised and offset for clarity. (b) FWHM of the fitted Voigt profiles, as a function of $\sqrt{P_{\rm yellow}}$. The solid line is a fit to the data (see text), from which we extract $a=18.6(1)\,\rm MHz/\sqrt{\rm nW}$ and $b=25.1(3)\,\rm MHz$. The transform limit, 7.6 MHz, arising from the excited state lifetime, $\tau_{\rm exc}=21\,\rm ns$, is shown as a dashed line.}
	\end{figure}

	In a second set of experiments, we study the linewidth of the observed optical transitions for various values of ${P_{\rm yellow}}$, see Fig.\,\ref{fig:broadening}(a). We observe a broadening of the lines with increasing power. As a result, in the high power regime the fine structure is no longer resolved. The extracted full width at half maximum (FWHM) is plotted in Fig.\,\ref{fig:broadening}(b) as a function of $\sqrt{P_{\rm yellow}}$ (i.e. $\propto$ optical Rabi frequency). For power broadening, a linear dependence on the Rabi frequency is expected, while at low powers the FWHM is limited by the intrinsic linewidth of the defect. In Fig.\,\ref{fig:broadening}(b) we fit the dependence under the assumption that the linewidth can be described by a convolution of a lifetime-limited Lorentzian profile ($f_{\rm L}=1/(2\pi\times\tau_{\rm exc})=7.6\,\rm MHz$) with Gaussian broadening terms. The resulting Voigt FWHM can be approximated by $f \approx 0.5446 f_{\rm L} + \sqrt{0.2166 f_{\rm L}^{2} + f_{\rm G}^{2}}$ \cite{olivero1977empirical}. The Gaussian component is given by the convolution of a power-dependent term from power-broadening, and a power-independent term arising from spectral diffusion: $f_{\rm G} = \sqrt{a^{2}P_{\rm yellow} + b^{2}}$. The fit shows good agreement with the observed behaviour. We find a spectral-diffusion-limited linewidth of $30.3(3)\,\rm MHz$, a factor of $4$ above the transform limit.
	
	\subsection{Extraction of the NV$^{0}$ fine structure parameters} 
	
	The obtained PL measurements carry information of the parameters of the ground and excited state Hamiltonians. Based on these measurements we develop a methodology to extract the NV$^0$ orbital g-factor $l$ and spin-orbit interaction parameter $\lambda$. 
	
	\begin{figure*}
		\includegraphics[width=1\linewidth]{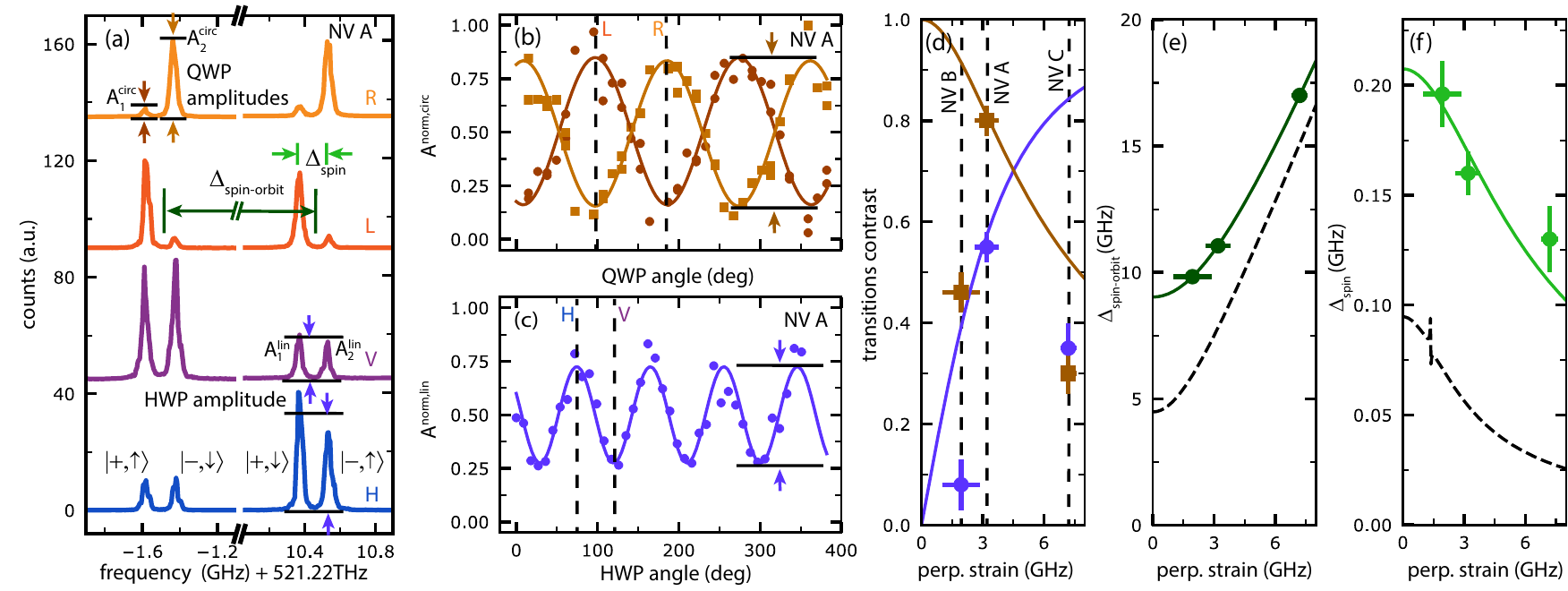}%
		\caption{\label{fig:finestructure} Extraction of finestructure constants. (a) PL spectra for four different excitation light polarisations (R, L, V, H) for NV A. (b) NV A: Example traces for extracted and normalized PL amplitudes $A^{\rm circ,norm}_1$ (circles) and $A^{\rm circ,norm}_2$ (squares) for varying angles of the quarter-wave plate (QWP). (c) Same as (b) but for $A^{\rm lin,norm}$ when varying the angle of the half-wave plate (HWP). The dashed lines in (b,c) indicate the angles for which the PL scans in (a) are taken. Solid lines show fits with a sine function. (d) Transition contrast for circular (`orbit contrast', squares) and linear (`spin-orbit contrast', circles) polarisation extracted from the normalized amplitudes, as indicated by the vertical arrows in (b,c), determined for three NV centers (dashed lines). (e,f) $\Delta_{\text{spin-orbit}}$ and $\Delta_{\mathrm{spin}}$ for the same NV centers as in (d). The data of (d-f) is simultaneously fitted against the Hamiltonians. The data is shown together with the best fit (solid lines) as a function of perpendicular strain $\epsilon_{\bot}$. Dashed lines in (e,f) show the expected transition energy splittings for the fine structure constants of Barson et al., see Ref.\,\cite{barson_fine_2019}.}
	\end{figure*}
	
	Beside the fine-structure constants, the PL spectrum of the NV$^0$ center depends on both magnetic field and strain (stress within the crystal and electric fields). These dependencies are captured within the NV$^0$ Hamiltonian of the $^2E$ ground state,  see Ref.\,\cite{barson_fine_2019}:
	\begin{equation}
	\begin{split}
	H  = & \: g\mu_{B}\hat{S}_{z}B_{z} + l\mu_{B}\hat{L}_{z}B_{z} + 2 \lambda \hat{L}_{z}\hat{S}_{z}\\ 
	& + \epsilon_{\bot} (\hat{L}_- + \hat{L}_+).
	\end{split}
	\end{equation}
	Here, $g$ is the spin g-factor, $\mu_{B}$ is the Bohr magneton, $\hat{L_{z}} = \sigma_{z}$ and $\hat{S_{z}} = \frac{1}{2}\sigma_{z}$ are the orbital and spin operators defined in terms of the Pauli matrix $\sigma_{z}$.
	The last term of the Hamiltonian shows the influence of perpendicular strain $\epsilon_{\bot}$, where $\hat{L}_{\pm}= \ket{\pm}\bra{\mp}$ with $\ket{\pm} = \mp(1/\sqrt{2}(\ket{X}\pm i\ket{Y}))$ and \{$\ket{X},\ket{Y}$\} is the basis for the strain eigenstates. Note that parallel strain is not included as it does not affect the relative energy of the ground state levels.
	For the excited state $^2A_2$, the Hamilonian reads as
	\begin{equation}
	H  =  g\mu_{B}\hat{S}_{z}B_{z}
	\end{equation}
	and does not show a dependency on strain.
	
	These two Hamiltonians lead to the energy level structure as presented in Fig.\,\ref{fig:levels_supp}. The four resulting transition frequencies are shown in the PL spectra in Fig.\,\ref{fig:finestructure}(a). The corresponding eigenstates in the ground state are indicated. Depending on the polarisation (circular right (R) or left (L); linear horizontal (H) or vertical (V)), the amplitude of the observed PL varies. From Voigt fits to the individual PL lines we extract the transition frequencies and PL amplitudes. From the transition frequencies, we determine the energy splitting, $\Delta_{\mathrm{spin}}$, between the two spin-states, $\ket{\uparrow}$ and $\ket{\downarrow}$, associated with each spin-orbit branch, and the energy splitting $\Delta_{\text{spin-orbit}}$ between the two spin-orbit branches.
	
	Further, the PL amplitudes can be directly related to the transition strength for a given polarization and transition. In the absence of strain, circularly-polarized transitions are expected, as one quantum of orbital angular momentum has to be transferred upon excitation. Accordingly, under such excitation, full PL contrast would be expected between the transitions within each spin-orbit branch (`orbit contrast'). However, under large strain, the ground state is better described within the strain eigenbasis~\cite{barson_fine_2019}, with associated linearly-polarised transitions. In this scenario, full contrast would instead be expected between the spin-orbit branches (`spin-orbit contrast'), whilst no `orbit contrast' would be expected within each branch. As a consequence the observed contrasts can be used to determine the strain. 
	
	To extract the contrast, we repeat PL scans of a spin-orbit branch for several angles of the quarter-wave plate (QWP) and half-wave plate (HWP). In the case of circular polarisation, we normalize each PL scan by the respective integrated total counts. The fitted amplitudes $A_1^{\rm circ}$ and $A_2^{\rm circ}$ for the transitions of two spin states are then normalized by the mean of the sum of the two amplitudes for all measured angles, $A_i^{\rm circ,norm}= A_i^{\rm circ}/\overline{A_1^{\rm circ}+A_2^{\rm circ}}$ with $i\in \{1,2\}$. For linear polarisation, for each angle we take the mean amplitude of the transitions within a spin-orbit branch, $A^{\rm lin} = (A^{\rm lin}_1+A^{\rm lin}_2)/2$. We then normalize each angle by the mean for all measured angles, $A^{\rm lin,norm}= A^{\rm lin}/\overline{A^{\rm lin}}$. Figure~\ref{fig:finestructure}(b,c) shows example plots for the resulting normalized amplitudes for varying circular and linear polarizations. Measurements on the respective other spin-orbit branch give similar results.
	
	As a next step we extract the transition contrast for circular polarizations (Fig.\,\ref{fig:finestructure}(b), `orbit contrast') and linear polarizations (Fig.\,\ref{fig:finestructure}(c), `spin-orbit contrast') from the amplitudes of fits with a sine function.
	These contrasts are plotted in Fig.\,\ref{fig:finestructure}(d) for three independently measured NV centers, NV A, NV B, and NV C. Further, we plot the mean transition energy splittings $\Delta_{\text{spin-orbit}}$ and $\Delta_{\mathrm{spin}}$ in Fig.\,\ref{fig:finestructure}(e) and (f).
	
	These three data sets are now fitted against the Hamiltonian with $l$, $\lambda$, and $\epsilon_{\bot}$ for each NV as free fitting parameters.
	Our results are summarized in Tab.\,\ref{tab:finestructure} (method 1). The fitted strain results are used to place the experimental data points in Fig.\,\ref{fig:finestructure}(d-f), see dashed lines in Fig.\,\ref{fig:finestructure}(d). Solid lines show the result of our fit.
	The transition energy splittings $\Delta_{\mathrm{spin}}$ and $\Delta_{\text{spin-orbit}}$ are well described by our theoretical model. 
	While a good match of the transition contrast for NV A is found, NV B and C show a discrepancy. A possible explanation could be non ideal polarization settings at the position of the NV for these data sets, resulting in less clean rotation around the Poincar\'e sphere, i.e. mixed circular and linear polarizations when rotating the QWP/HWP. A mixed polarization leads to both reduced `orbit contrast' and reduced `spin-orbit contrast'.
	\begin{table}[h!]
		\begin{center}
			\begin{tabular}{|c||c c c c c||} 
				\hline
				method & $\epsilon_{\bot}^{\rm A}$ (GHz) & $\epsilon_{\bot}^{\rm B}$ (GHz) & $\epsilon_{\bot}^{\rm C}$ (GHz) & $l$ & $\lambda$ (GHz) \\ [0.5ex] 
				\hline\hline
				1 & 1.9(9) & 3.2(6) & 7.2(4) & 0.040(8) & 4.5(4)\\ [1ex] 
				\hline
				2 & 1.05 & 4.15 & 4.35 & 0.037(14) & 5.2(4)\\ [1ex] 
				\hline
			\end{tabular}
			\caption{Ground state Hamiltonian parameters extracted from a fit to our experimental data for the three NV centers NV A, NV B, and NV C. For method 1 the measured contrasts of NV$^0$ is used as an input to fit the strain values, while method 2 uses the strains extracted from NV$^-$ as fixed parameters.}
			\label{tab:finestructure}
		\end{center}
	\end{table}
	
	As a second method we repeat our fitting procedure, but this time fixing the individual strain values to $\epsilon_{\rm NV^-}$ of each NV, obtained from NV$^-$, see Fig.\,\ref{fig:levels_supp}. While it is not known how the NV$^-$ strain translates to strain in NV$^0$, a correlation is expected.
	From the fit we obtain $l$ and $\lambda$, see Tab.\,\ref{tab:finestructure} (method 2).
	Within error the two methods give the same values. In the main text we report the mean of the two values: $l_{\mathrm{mean}}=0.039(11)$, $\lambda_{\mathrm{mean}}=4.9(4)\,\mathrm{GHz}$.
	
	We now compare our data to the transition frequencies calculated using the fine structure parameters of Barson et al.\,\cite{barson_fine_2019}, $l_{\mathrm{lit}}=0.0186(5)$, $\lambda_{\mathrm{lit}}=2.24(5)\,\mathrm{GHz}$, (dashed lines, Fig.\,\ref{fig:finestructure}(e,f)). 
	Strikingly, with these parameters, our data for $\Delta_{\rm spin}$ cannot be reproduced for \emph{any} strain value, strongly indicating that the discrepancy in fine structure parameters cannot be explained by systematic errors in our method to extract the strain of the NV.
	
	We note that Barson et al.\, have used NV ensemble magnetic-circular dichroism measurements while we here observe PL of single NV centers.
	
	\subsection{Charge-resonance check}
\begin{figure*}
		\includegraphics[width=1\linewidth]{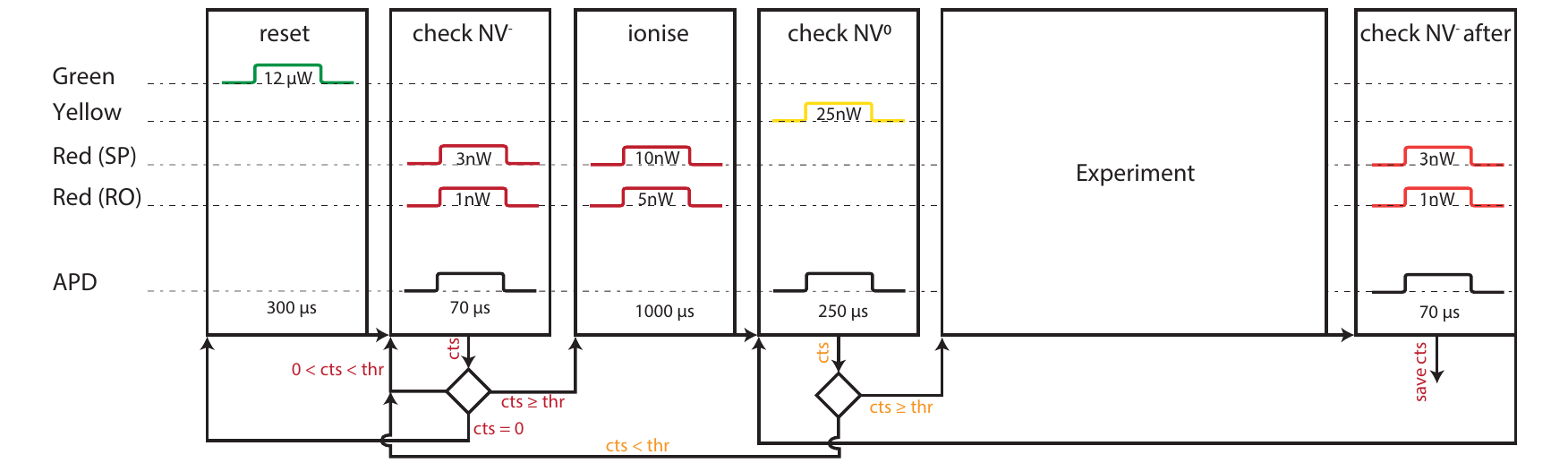}%
		\caption{\label{fig:antiCR} NV$^-$ and NV$^0$ CR check. The application of laser pulses, counting of PSB photons, and decisions depending if the measured counts (cts) have passed the preset thresholds (thr) are all performed via a micro-processor (ADWIN).}
	\end{figure*}

	As described in the main text, one of the key components of the experiments in this work is the introduction of a charge-resonance (CR) check for NV$^{0}$. This check allows heralded preparation of the NV in the neutral charge state, while also preparing the red lasers on resonance with the NV$^{-}$ optical transitions, and the yellow laser on resonance with one of the four NV$^{0}$ transitions. The full procedure is shown in Fig.\,\ref{fig:antiCR} and outlined in the following.
	
	We first prepare the negatively-charged state. A strong green pulse ($12\,\rm \mu W$) is applied for $300\,\rm \mu s$ in order to prepare the system (`reset'). Next, we simultaneously apply a combination of the red RO ($1\,\rm nW$) and SP ($3\,\rm nW$) light for a duration of $70\,\rm \mu s$, during which time we integrate all single-photon counts incident on an avalanche photo-diode (APD) (`check NV$^{-}$'). If a set photon-count threshold is exceeded, we have high confidence that the red lasers are well on resonance with their associated transitions and that the NV is in NV$^{-}$, and proceed to the next step. If the count is below the threshold, but above zero, then it is assumed that the NV is in the negative charge state and close to resonance, but not yet in a satisfactory regime. In this case, the red check is repeated until the threshold is passed. In the case that any red check produces a zero photon-count, the green pulse (`reset') is reapplied to reset the charge state or to induce significant spectral diffusion bringing the NV$^-$ transitions back in resonance with the red lasers. We note that the low powers used for resonant excitation itself cause minimal spectral diffusion.
	
	After the NV$^{-}$ check, an ionization pulse is applied to prepare NV$^{0}$ (`ionise'). Here, we apply $5\,\rm nW$ ($10\,\rm nW$) of RO (SP) light respectively, for a total duration of $1\,\rm ms$. While the ionisation probability is low ($\sim 2\%$), the chosen powers ensure that spectral diffusion is minimal.
	
	To verify that the NV has been successfully transferred to the neutral charge state and to confirm that the yellow light is on resonance with a single transition we apply $25\,\rm nW$ of yellow light for $250\,\rm \mu s$ (`check NV$^{0}$'). In this check, we either exceed the threshold, in which case we proceed to the main experiment, or we return to the `check NV$^{-}$' step.  The counts of each successful `check NV$^{0}$' step are saved. 
	
	In our experiments, we use a single yellow laser only. We note that polarization of this laser affects the spin-state prepared after the CR check. When exciting with linear polarization, one of the two spin states is prepared in each experimental repetition. Which spin state is prepared may vary due to spectral diffusion between repetitions in combination with the close spectral vicinity between the spin states (see Fig.\,\ref{fig:finestructure}(a)).
	However, when exciting with circular polarization, a single NV$^{0}$ spin-state is selectively addressed and heralded throughout: the probability to false-herald a non-targeted transition is negligible. While the laser frequency corresponds to a transition associated with a specific spin-orbit state, the check heralds a mixed orbital state as it takes significantly longer than the orbital relaxation time. We note that a general (not spin-selective) resonance check with higher efficiency could be achieved by adding a second laser to address a NV$^{0}$ transition corresponding to the opposite spin-state. 
	
	Finally, after the experiment has been completed, we perform the `check NV$^{-}$ after' step, which can be used to detect transfer from NV$^{0}$ to NV$^{-}$ during the experiment. This is enabled by the fact that transfer between the NV charge states induces minimal spectral diffusion, as witnessed by our recharging data (see corresponding supplementary section). 
	
	We then return to the `check NV$^{0}$' step due to two reasons. First, the detected counts can be used to readout the NV$^{0}$ spin state in the case recharging has not happened. Second, for most experimental repetitions we remain in NV$^{0}$, leaving a high probability that the `check NV$^{0}$' step is passed again. In such cases, it is not necessary to repeat the `check NV$^{-}$' step, reducing measurement overhead time.
	
	\subsection{Lindblad master equation simulations}
	
	In order to capture the full dynamics of the NV$^0$ system we develop a theoretical model based on the Lindblad master equation.
	In Eq.\,\ref{eq:initstate}, we describe the initial (mixed) state of the system, whose populations are parameterized by probabilities $p_i$ for $i \in \{0,..,5\}$. Eq.\,\ref{eq:master} describes the non-unitary time evolution of that state, $\rho(t)$:
	
	\begin{widetext}
		\begin{equation}
		\label{eq:initstate}
		\rho_0 =  p_0\ket{+,\downarrow}\bra{+,\downarrow}+p_1\ket{-,\uparrow}\bra{-,\uparrow}+p_2\ket{-,\downarrow}\bra{-,\downarrow}+
		p_3\ket{+,\uparrow}\bra{+,\uparrow}+p_4\ket{\rm 0,\downarrow}\bra{\rm 0,\downarrow}+p_5\ket{\rm 0,\uparrow}\bra{\rm 0,\uparrow}
		\end{equation}
		\begin{equation}
		\label{eq:master}
		\dot{\rho}(t) = -\frac{i}{\hbar}[H(t),\rho(t)] + \sum_n C_n\rho(t)C_n^\dag  - \frac{1}{2}[\rho(t)C_n^\dag C_n - C_n^\dag C_n\rho(t)].
		\end{equation}
	\end{widetext}
	
	Here, $H(t)$ is the time-dependent Hamiltonian, $\hbar$ the Planck constant and $C_n=\sqrt{\gamma}A_n$ the collapse operator capturing relaxation processes. $A_n$ is the coupling operator and $\gamma=1/\tau_{\rm relax}$ the corresponding decay rate with relaxation time constant $\tau_{\rm relax}$.
	
\begin{figure}
		\includegraphics[width=1\linewidth]{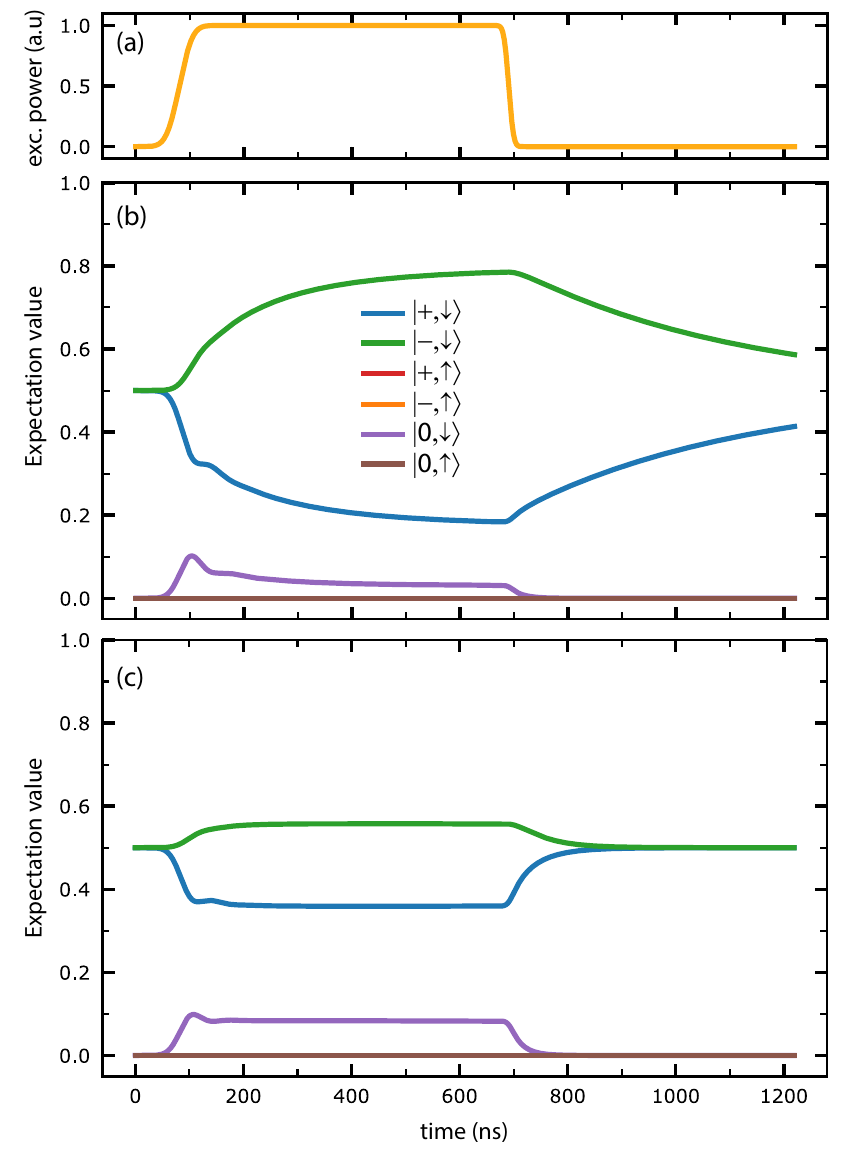}
		\caption{\label{fig:Master} Master equation simulations. (a) Shape of the experimental yellow pulse with measured AOM rise and fall times of $30(5)$ and $7(1)\,\mathrm{ns}$, respectively. (b,c) Simulated expectation value for the relevant NV$^{0}$ levels (inset) under resonant excitation of $\ket{+,\downarrow}$ for $\tau_{\rm orbit}=425\,\rm ns$ ($T=4.65\,\mathrm{K}$) (b) and $\tau_{\rm orbit}=50\,\rm ns$ ($T=10.1\,\mathrm{K}$) (c). For the chosen power, population of the $\ket{\uparrow}$ spin levels remains zero throughout the simulation.
		}
	\end{figure}

	We now show an example case for dynamics under optical pumping (see Fig.\,2 of the main text). In the experiment, we herald a specific spin-orbit state via the `check NV$0$' step, simultaneously bringing the yellow laser on resonance with the respective transition. As an example, we can herald the $\ket{+,\downarrow}$ state. However, due to orbital relaxation dynamics being much faster then the duration of the state check we effectively herald an orbitally mixed $\ket{\downarrow}$ state, i.e.
	\begin{equation}
	\label{eq:startstate}
	\rho_0 =  \frac{1}{2}\left(\ket{+,\downarrow}\bra{+,\downarrow}+\ket{-,\downarrow}\bra{-,\downarrow}\right ).
	\end{equation}
	Under optical pumping, we can describe the (rotating frame) system Hamiltonian, $H(t)$, and the total collapse operator, $\sum_n C_n$ as:
	\begin{widetext}
		\begin{equation}
		\label{eq:hamiltonian_supp}
		H(t) = 2\pi \hspace{10pt}
		\begin{blockarray}{ccccccc}
		\ket{+,\downarrow} & \ket{-,\uparrow} & \ket{-,\downarrow} & \ket{+,\uparrow} & \ket{\rm 0,\downarrow} & \ket{\rm 0,\uparrow} \\
		\begin{block}{(cccccc)c}
		0 & 0 & 0 & 0 & \frac{\Omega(t)}{2} & 0 &\:\bra{+,\downarrow}\\
		0 & 0 & 0 & 0 & 0 & \frac{\Omega(t)}{2} & \:\bra{-,\uparrow}\\
		0 & 0 & 0 & 0 & 0 & 0 & \:\bra{-,\downarrow}\\
		0 & 0 & 0 & 0 & 0 & 0 & \:\bra{+,\uparrow}\\
		\frac{\Omega(t)}{2} & 0 & 0 & 0 & \delta & 0 & \:\bra{\rm 0,\downarrow}\\
		0 & \frac{\Omega(t)}{2} & 0 & 0 & 0 & \Delta+\delta & \:\bra{\rm 0,\uparrow}\\
		\end{block}
		\end{blockarray}
		\end{equation} 
		\begin{equation}
		\label{eq:collapse}
		\sum_n C_n = \frac{1}{\sqrt{2}}\hspace{10pt}
		\begin{blockarray}{ccccccc}
		\ket{+,\downarrow} & \ket{-,\uparrow} & \ket{-,\downarrow} & \ket{+,\uparrow} & \ket{\rm 0,\downarrow} & \ket{\rm 0,\uparrow} \\
		\begin{block}{(cccccc)c}
		0 & 0 & \sqrt{\frac{1}{\tau_{\rm orbit}}} & \sqrt{\frac{1}{\tau_{\rm spin}}} & \sqrt{\frac{1}{\tau_{\rm exc}}} & 0 & \:\bra{+,\downarrow}\\
		0 & 0 & \sqrt{\frac{1}{\tau_{\rm spin}}} & \sqrt{\frac{1}{\tau_{\rm orbit}}} & 0 & \sqrt{\frac{1}{\tau_{\rm exc}}} & \:\bra{-,\uparrow}\\
		\sqrt{\frac{1}{\tau_{\rm orbit}}} & \sqrt{\frac{1}{\tau_{\rm spin}}} & 0 & 0 & \sqrt{\frac{1}{\tau_{\rm exc}}} & 0 & \:\bra{-,\downarrow}\\
		\sqrt{\frac{1}{\tau_{\rm spin}}} & \sqrt{\frac{1}{\tau_{\rm orbit}}} & 0 & 0 & 0 & \sqrt{\frac{1}{\tau_{\rm exc}}} & \:\bra{+,\uparrow}\\
		0 & 0 & 0 & 0 & 0 & \sqrt{\frac{1}{\tau_{\rm exc,spin}}} & \:\bra{\rm 0,\downarrow}\\
		0 & 0 & 0 & 0 & \sqrt{\frac{1}{\tau_{\rm exc,spin}}} & 0 & \:\bra{\rm 0,\uparrow}\\
		\end{block}
		\end{blockarray}
		\end{equation}
	\end{widetext}
	The diagonal elements in the Hamiltonian of Eq.\,\ref{eq:hamiltonian_supp} correspond to the detuning of each transition with respect to the laser frequency, while the off-diagonal elements enable Rabi driving between respective levels. In our example, the laser is on resonance with the $\ket{+,\downarrow}\xleftrightarrow{}\ket{0,\downarrow}$ transition. The transition $\ket{-,\uparrow}\xleftrightarrow{}\ket{0,\uparrow}$ has a detuning of $\Delta=160\,\rm MHz$. An additional detuning, $\delta$, is randomly sampled from a Gaussian distribution with FWHM $=2\pi\times 20\,\mathrm{MHz}$, to account for the effects of imperfect laser resonance checks or small spectral diffusion. The Rabi frequency is given by $\Omega(t)=\alpha\sqrt{P(t)}$, with $\alpha$ being a proportionality factor and $P(t)$ the time dependent laser power.
	Note that we neglect driving of the (far-detuned) upper spin-orbit states and thus omit the corresponding terms. 
	In our simulations we implement $P(t)$ with rise/fall times as independently measured in our experiment, see Fig.\,\ref{fig:Master}(a). 
	The elements in Eq.\,\ref{eq:collapse} correspond to relaxation processes between certain levels, see Fig.\,\ref{fig:levels_supp}. Here we use the respective timescales as extracted within the main text. 
	
	In Fig.\,\ref{fig:Master}(b-c) we plot the simulated expectations values for two different temperature scenarios, both for a driving power of $5\,\mathrm{nW}$. In both cases we observe that the $\ket{+,\downarrow}$ state is depopulated, while population in the opposite orbital state, $\ket{-,\downarrow}$, grows via the excited state, $\ket{\rm 0,\downarrow}$. Initially, a damped Rabi oscillation between $\ket{+,\downarrow}$ and $\ket{0,\downarrow}$ is observed, before a steady state population is reached. In the case of lower temperature and hence a slower orbital relaxation constant, $\tau_{\mathrm{orbit}}$, a stronger orbital pumping is observed. After the laser light is switched off, the excited state decays with $\tau_{\mathrm{exc}}$, while the two orbital states relax back to an equal population with the time constant $\tau_{\mathrm{orbit}}$.
	
	In Fig.\,2 of the main text we plot the expectation value of the excited state $\ket{\rm 0,\downarrow}$, and find an excellent agreement with the experimental fluorescence counts. This confirms the accuracy of our theoretical model, and shows that the observed NV$^0$ dynamics are well understood.
	We note that the timescales for spin relaxation, $\tau_{\mathrm{spin}}$, and spin pumping, $\tau_{\mathrm{exc,spin}}$, are much longer than the pumping duration, and hence are not observed in this set of experiments.
	
	\subsection{Recharging dynamics}
	
	An important feature of our experiments is the ability to switch between the NV$^{-}$ and NV$^{0}$ charge states through resonant excitation. Ionisation (NV$^{-}\xrightarrow{}$ NV$^{0}$) under resonant excitation has previously been experimentally studied ~\cite{robledo_control_2010,siyushev_optically_2013}, and an ionisation mechanism was proposed that combines a two-photon and an Auger process~\cite{siyushev_optically_2013}. 
	While a mechanism for the recharging process (NV$^{0}\xrightarrow{}$ NV$^{-}$) was also proposed, the power dependence under resonant excitation has not been measured \cite{siyushev_optically_2013}. Here we present such measurements.
	
\begin{figure}
		\includegraphics[width=1\linewidth]{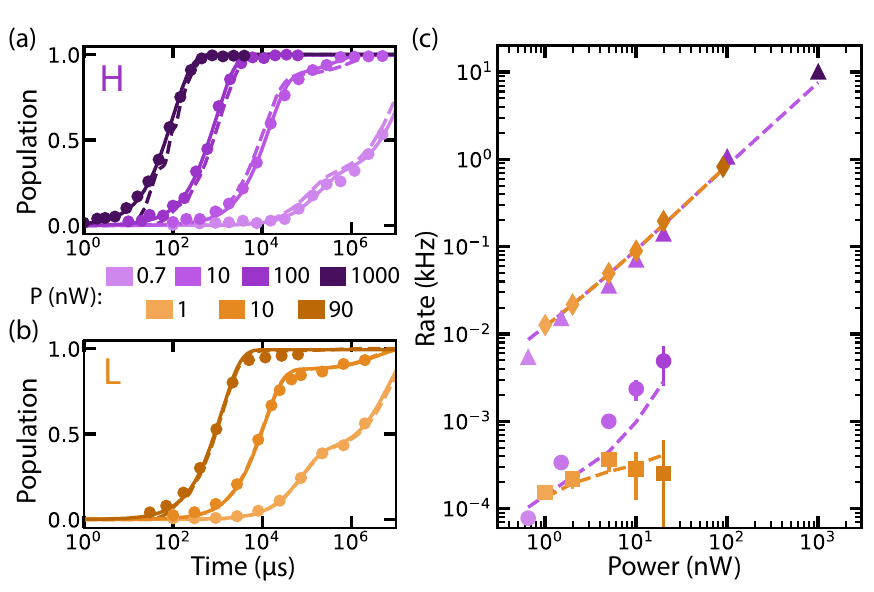}%
		\caption{\label{fig:Recharging} NV$^{-}$ population as a function of recharging time after initialisation into NV$^{0}$, for horizontal (a), and left circular (b) polarisation, respectively. Solid lines are fits following a double-exponential growth function, while dashed lines are populations obtained by Master equation simulations (see text). (c) Fitted fast and slow timescales as a function of the recharging power. Markers correspond to measured data for linear (triangles and circles) and circular (diamonds and squares) polarisation, while the dashed lines connect the values obtained from numerical simulations.}
	\end{figure}

	We first herald the defect in the NV$^{0}$ state, see `check NV$^{0}$' step in the previous section, with the yellow laser resonant to a single transition in the lower spin-orbit branch. We then apply yellow recharging light for a certain time, before we read-out the NV$^{-}$ population via the `check NV$^{-}$ after' step.
	In Fig.\,\ref{fig:Recharging}(a) (Fig.\,\ref{fig:Recharging}(b)) we plot the NV$^{-}$ population as a function of recharging time for linear (circular) polarization, for a few selected powers. For all powers we observe a growth in NV$^{-}$ population that eventually approaches unity both under high- and low-power excitation. This important observation shows both that high-fidelity switching can be achieved, and that the red read-out is robust: spectral diffusion remains minimal even for second long experiments. Additionally, the ability to reach near-unity NV$^{-}$ population within a few-hundred $\mathrm{\mu s}$ suggests that significant population is not trapped in the $^{4}A_{2}$ level of NV$^{0}$. Possible explanations for this observation are that the inter-system crossing rate is small, the $^{4}A_{2}$ lifetime is short, or the recharging process can also occur from the $^{4}A_{2}$ level itself.
	
	For a range of recharging powers, we perform sweeps as exemplified in Figure \ref{fig:Recharging}(a) and (b), and fit with growth functions. For comparison, we additionally perform Master equation simulations, for which we can fit the same growth functions. In the simulations the NV$^-$ state is implemented as a dark state that can be populated via an excitation of the $^2A_2$ NV$^0$ state to the conduction band. Orbital dynamics are neglected as their corresponding timescales are much faster then the observed timescales for recharging. Further, in order to perform the simulations in reasonable computational time, the recharging rate, spin-pumping rate, and spin-relaxation rates are re-scaled by a fixed factor of four orders of magnitude. Remarkably, our simulations show both qualitative and quantitative agreement with the experimental data. At very short times small deviations arise as the (unmodified) excited-state lifetime becomes comparable to the re-scaled recharging process.
	
	We generally anticipate two recharging timescales. A fast timescale, $\tau_{\rm fast}$, is associated with resonant recharging. A slow timescale, $\tau_{\rm slow}$,  arises from NV$^0$ spin-pumping and spin-relaxation causing the driven transition to become dark, i.e. via a spin flip $\ket{\downarrow}\xleftrightarrow{}\ket{\uparrow}$. Recharging from the dark state occurs due to a combination of off-resonant excitation and relaxation back to the resonantly driven spin-state. Note that a slow timescale is expected for both linear and circular polarisation. The initial CR check prepares a single spin-state, and at low powers, both polarisations perform spin-selective driving as power-broadening is significantly less than the detuning between the $\ket{\uparrow}$ and $\ket{\downarrow}$ optical transitions. 
	The behaviour can thus be described by a double-exponential growth function: $f(t_{\rm recharge}) = A e^{-t_{\rm recharge}/\tau_{\rm fast}} + (1-A)e^{-t_{\rm recharge}/\tau_{\rm slow}}$. Where it is not possible to fit a second timescale due to the dominance of the fast timescale (high powers), a single exponential growth function is used, $f(t_{\rm recharge}) = A e^{-t_{\rm recharge}/\tau_{\rm fast}}$.
	
	In Fig.\,\ref{fig:Recharging}(c), we plot all of the fitted rates. Two key features are apparent. First, for both linear and circular polarisation, the fast timescales broadly overlap and follow a linear power-dependence, as expected for a two-photon process for which the first step is resonant and in saturation. From a linear fit we extract a recharging rate of 9.3(6) Hz/nW. Second, the slow timescale is seen to be significantly faster for linear than for circular polarisation. This difference arises from the fact that off-resonant excitation is strongly suppressed by circular polarisation, preventing both spin-pumping back to the resonant transition and off-resonant recharging. 
	The extracted rates from the Master equation simulations are in good agreement with our experimental data both for the fast and the slow timescales.
	
	Importantly, this characterisation can be used to determine the frequency of charge-resonance checks required in each experiment to ensure that the NV$^{0}$ population remains high.
	
	\subsection{Pump-probe spectroscopy}
	
	Here, we describe the procedures followed for the pump-probe spectroscopy measurements, Fig.\,3 in the main text.
	
	For each set temperature of the cryostat, we measure a series of time traces as exemplified in Fig.\,3(a) of the main text. For each time trace, we integrate the total photon counts during the first 40 ns after opening the AOM for the probe and pump measurements, respectively, and take their ratio. We then fit the recovery behaviour against the delay time, $t_{\rm delay}$, following the function $f(t_{\rm delay}) = a + A (1-e^{-(t_{\rm delay}-t_0)/T})$, where $a$ is the steady-state fluorescence under pumping, $A$ is the peak fluorescence of the mixed state, $t_0$ is the time at which the pump is turned off, and $T$ is the characteristic recovery time. 
	
	At higher temperatures, a challenge for these measurements arises as the required time resolution exceeds that of the AOM rise- and fall-times, see Fig.\,\ref{fig:Master}(a), and hence the optical pulse is not turned off for the entirety of the set $\mathrm{t_{delay}}$. We correct for this systematic error by calibrating the delay times as those for which the optical pulse falls below 90\% of the full amplitude, measured using a fast photodiode. However, we note that some corrections may remain, which may lead to an underestimate of the orbital relaxation rates in those measurements. In future work, such measurements could be improved using a fast electro-optic modulator to gate the pulses. 
	
	\subsection{Rate Equations}
	
	In order to extract the timescales for recharging and spin-pumping, as shown in Fig.\,4 of the main text, we develop a three-level model for which we can derive analytic solutions for use in a fitting routine.
	
	We start with the scenario of Fig.\,4(c). Given the relatively slow timescales under consideration (for 5 nW of yellow excitation), we choose to neglect the orbital basis (which is effectively mixed, see Eq.\,\ref{eq:startstate}). This leaves three levels, $\ket{\uparrow}$, $\ket{\downarrow}$ and NV$^{-}$, which we denote as $U(t)$, $D(t)$, $N(t)$. We consider the processes of resonant recharging ($r$, from $D \xrightarrow{} N$), resonant spin-pumping ($p$, from $D \xrightarrow{} U$) and spin-relaxation ($s$, from $D \xleftrightarrow{} U$), and neglect off-resonant recharging (from $U \xrightarrow{} N$), off-resonant spin-pumping (from $U \xrightarrow{} D$), and ionisation (from $N \xrightarrow{} D,U$), which are all expected to have negligible rates in this parameter regime.  The dynamics between these levels are thus described by the following coupled equations:
	
	\begin{gather}
	N(t) + D(t) + U(t) = 1 \\
	\frac{dN(t)}{dt} = r \, D(t) \\
	\frac{dD(t)}{dt} = -(r+s+p) \, D(t) + s \, U(t) \\
	\frac{dU(t)}{dt} = +(s+p) \, D(t) - s \, U(t) 
	\end{gather}
	
	We impose initial conditions
	\begin{gather}
	D(0) = c_{1} \\
	U(0) = c_{2}\\
	N(0) = 1 - c_{1} - c_{2}
	\end{gather}
	and derive analytic solutions to these equations (solutions available in code form upon request), which we then use as fitting functions for the measured populations. The spin relaxation time is fixed to the independently measured time of 1.51(1) s. All other parameters are free in the optimisation, leading to the following best-fit values:
	\begin{table}[h!]
		\begin{center}
			\begin{tabular}{|c || c c c c|}  \hline
				$\tau_{\rm spin}$ (s) & $\tau_{\rm recharge}$ (s) & $\tau_{\rm pump}$ (s) & $c_{1}$ & $c_{2}$ \\ [0.5ex] \hline
				1.51 & 0.027(1) & 0.090(4) & 0.960(6) & 0.012(5)\\ [1ex] 
				\hline
			\end{tabular}
		\end{center}
	\end{table}
	
	Here, $\tau_{\rm spin} = 1/s$, $\tau_{\rm recharge} = 1/r$, $\tau_{\rm pump} = 1/p$, and $c_1$ and $c_2$ are the respective populations in $\ket{\downarrow}$ and $\ket{\uparrow}$ after initialisation.
	
	We now move to the scenario shown in Fig.\,4(d). In this case, we stroboscopically interleave periods of $5\,\rm nW$ yellow excitation with periods of $500\,\rm nW$ of red NV$^-$ spin-pumping excitation coupled with hard microwave $\pi$-pulses regularly spaced by $1.25\,\rm \mu s$. To simplify this into a rate-equation model, we make the following assumptions (along with those already made for the previous case). First, we combine the stroboscopic driving into continuous driving with averaged resonant recharging ($r$, from $D \xrightarrow{} N$) and ionisation ($i$, from $N \xrightarrow{} D, U$) rates. We assume that charge conversion from NV$^{-}$ to NV$^{0}$ results in a completely mixed NV$^{0}$ spin state: that is, the rate, $i$, equally couples to both $D$ and $U$. We note that analytic solutions incorporating an asymmetry in these couplings were derived, but that any asymmetry could not be constrained by the fitting procedure. Finally, we assume that the microwaves only couple the (not-described) spin levels of NV$^{-}$ --- preventing spin-pumping in that charge state --- and so only influence the rate $i$. Under this model, we arrive at the following set of coupled equations:
	
	\begin{gather}
	N(t) + D(t) + U(t) = 1 \\
	\frac{dN(t)}{dt} = r \, D(t) - i \, N(t)\\
	\frac{dD(t)}{dt} = -(r+s+p) \, D(t) + s \, U(t) + \frac{i}{2} \, N(t)\\
	\frac{dU(t)}{dt} = +(s+p) \, D(t) - s \, U(t) + \frac{i}{2} \, N(t)
	\end{gather}
	
	We again impose initial conditions
	\begin{gather}
	D(0) = c_{1} \\
	U(0) = c_{2} \\
	N(0) = 1 - c_{1} - c_{2}
	\end{gather}
	and derive analytic solutions to these equations (solutions available in code form upon request), which we then use as fitting functions for the measured populations. For this scenario, we fix $r$, $p$, $c_{1}$ and $c_{2}$ to the values obtained from the fit to the previous case, as the initialisation step and yellow excitation parameters are unchanged. The spin-relaxation rate, $s$, is fixed to twice its previous value ($t_{\rm spin} = 1.51/2$), as the time-axis on this plot is the total yellow excitation time, which is half of the total sequence time. This leaves only the ionisation rate unfixed, for which we obtain a fitted value of $\tau_{\rm ion} = 1/i = 0.018(4)$ s. While the behaviour at short times is well described by the model, the long-time behaviour is not well captured.  If all fit parameters are unconstrained, we find that the long-time behaviour is well described, and still obtain agreement to within 50\% for all values, aside from for $\tau_{\rm spin}$, for which we now fit $0.14(1)$ s. This could indicate that the presence of red excitation and/or strong microwave driving induces an additional NV$^{0}$ spin-relaxation mechanism.
	
	\subsection{Spin cyclicity}
	
	In order to extract the cyclicity of the $\ket{\downarrow}$ transition under resonant excitation as reported in the main text we use the following procedure. 
	Here, we are interested on how many photons are scattered before the $\ket{\downarrow}$ spin state is left, i.e. before leaving the three-level system $\{\ket{+,\downarrow}, \ket{-,\downarrow}, \ket{0,\downarrow}\}$.
	The experiment starts when $\ket{+,\downarrow}$ is heralded, which can then be excited to $\ket{0,\downarrow}$ (for $P_{\rm yellow} = 5\,$nW a Rabi flop takes about $42\,$ns). From here the electron decays ($\tau_{\rm exc}=22(1)\,$ns) either to $\ket{+,\downarrow}$ or to $\ket{-,\downarrow}$, each with a 50\% probability.
	If it decays to $\ket{+,\downarrow}$ it can immediately be excited again, while $\ket{-,\downarrow}$ corresponds to a dark state that can only relax back to $\ket{+,\downarrow}$ via orbital relaxation ($\tau_{\rm orbit}=0.43(6)$ $\mu$s).
	This process continues until either a charge- or spin-flip event happens. From Fig.\,4(c) we have extracted characteristic timescales of $27(1)\,\mathrm{ms}$ ($90(4)\,\mathrm{ms}$) for the charge conversion (spin pumping) process. Within this regime the cyclicity is limited by charge conversion, for which we calculate $0.98(7)\times 10^5$ scattered photons before leaving the $\ket{\downarrow}$ manifold. 
	
	We note that at lower powers the recharging process, which requires a two-photon excitation, is expected to be suppressed. Here, the cyclicity will then be limited by a spin-flip process, enabling $3.2(2)\times 10^5$ cycles.
	
	\subsection{Readout fidelity}
	In the main text, we report single-shot read-out (RO) of the NV$^{0}$ spin state with fidelity, $F_{\rm RO} \geq 98.2(9)\%$. Here we outline the characterisation procedure. 
	
	The readout fidelity is defined by the relation $F_{\mathrm{RO}} = \frac{1}{2}(F_{\downarrow|\downarrow} + F_{\uparrow|\uparrow})$, where $F_{i|j}$ is the probability to assign the spin state $\ket{i}$ after preparing $\ket{j}$. Note that this assumes that each spin state can be initialised perfectly. Imperfect initialisation will lead to a decrease in the achievable fidelity: the calculated RO fidelities are thus a lower bound.
	
	In the presented experiments, we use a single yellow laser for both state initialisation and readout. As this means that we are only able to herald the $\ket{\downarrow}$ state with high fidelity, we use the intrinsic spin relaxation process of NV$^{0}$ to prepare a mixed state, which can also be used to calculate the fidelity for $\ket{\uparrow}$ read-out. The histograms presented in Fig.\,4(a) of the main text are used for this calculation. 
	
	First, to calculate $F_{\downarrow|\downarrow}$ and $F_{\uparrow|\downarrow}$, we herald $\ket{\downarrow}$ using a threshold of 25 photons for the `check NV$^0$' step of Fig.\,\ref{fig:antiCR}. After a brief delay ($0.1\,\rm ms$), we check for any residual population in NV$^{-}$ (`check NV$^-$ after'), which is discarded. We then perform the `check NV$^0$' step again to read out the spin population. From 3000 experimental shots, we discard 52 cases, corresponding to an NV$^{-}$ population of 1.7(2)\%. Of the remainder, 98.4(2)\% of cases match or exceed the chosen threshold of 5 photons, see main text. We thus have:  $F_{\downarrow|\downarrow}$ = 98.4(2)\%, and $F_{\uparrow|\downarrow}$ = 1.6(2)\%. We note that the initialisation fidelity of the $\ket{\downarrow}$ heralding step ($25\,\rm nW$ for $250\,\rm \mu s$) is likely limited by spin pumping to $\ket{\uparrow}$. From independent spin-pumping measurements we expect a reduction of that population by $\sim 0.8\%$ over the 250 $\mu$s, though this is partially mitigated by the initialisation threshold of 25 photons.
	
	To calculate $F_{\uparrow|\uparrow}$ and $F_{\downarrow|\uparrow}$, we repeat the procedure, but now wait for 10 s between the first `check NV$^0$' step and the `check NV$^-$ after' step, preparing the mixed state. We anticipate a state preparation infidelity of $<$ 0.2\% arising from the finite waiting time.  
	
	After preparation of the fully mixed state, the RO outcomes are described by:
	
	\begin{gather}
	\label{eq:ssro1}
	F_\downarrow =  \frac{1}{2}(F_{\downarrow|\uparrow}+ F_{\downarrow|\downarrow}) \\
	F_\uparrow =  \frac{1}{2}(F_{\uparrow|\uparrow} + F_{\uparrow|\downarrow})
	\end{gather}
	
	where $F_\downarrow$ ($F_\uparrow$) is the probability to obtain $\geq$ 5 photons ($<$ 5 photons). In this experiment, we discard 35 cases attributed to NV$^{-}$, estimating the NV$^{-}$ population to be 1.2(2)\%. From the remaining cases, we find $F_\uparrow$ = 49.8(9)\% and $F_\downarrow$ = 50.2(9)\%. Using the values previously obtained for $F_{\downarrow|\downarrow}$ and $F_{\uparrow|\downarrow}$, we obtain $F_{\uparrow|\uparrow}$ = 98(2)\%, $F_{\downarrow|\uparrow}$ = 2(2)\%.
	
	Combining the results we arrive at a single-shot read-out fidelity, $F_{RO} \geq 98.2(9)\%$.

\end{document}